\documentclass{emulateapj}
\usepackage{natbib}
\usepackage{apjfonts}
\newcommand{\CII}{\ion{C}{2}}
\newcommand{\HI}{\ion{H}{1}}
\newcommand{\HII}{\ion{H}{2}}
\newcommand{\Msol}{$M_\odot$}
\newcommand{\kms}{\mbox{km~s$^{-1}$}}
\newcommand{\mjybm}{\mbox{mJy~bm$^{-1}$}}

\newcommand{\tsyse}{\mbox{$T_{\rm sys}^*$}}
\newcommand{\tmb}{\mbox{$T_{\rm MB}$}}

\slugcomment{Accepted for publication in ApJ, 7 April 2006}

\shorttitle{Dense Molecular Gas in the LMC}

\begin{document}

\title{Synthesis Imaging of Dense Molecular Gas in the\\
  N113 \HII\ Region of the Large Magellanic Cloud}

\author{Tony Wong\altaffilmark{1}, John B. Whiteoak, \& J\"urgen Ott}
\affil{CSIRO Australia Telescope National Facility, PO Box 76, Epping
  NSW 1710, Australia} 
\email{Tony.Wong@csiro.au}
\author{Yi-nan Chin}
\affil{Department of Physics, Tamkang University, 25137 Tamsui, Taipei, Taiwan}
\and
\author{Maria R. Cunningham}
\affil{School of Physics, University of New South Wales, Sydney NSW
  2052, Australia}
\altaffiltext{1}{also at School of Physics, University of New South
  Wales, Sydney NSW 2052, Australia}

\begin{abstract}

We present aperture synthesis imaging of dense molecular gas in the
Large Magellanic Cloud, taken with the prototype millimeter receivers
of the Australia Telescope Compact Array (ATCA).  Our observations of the
N113 \HII\ region reveal a condensation with a size of $\sim$6\arcsec\
(1.5 pc) FWHM, detected strongly in the 1--0 lines of HCO$^+$, HCN and
HNC, and weakly in C$_2$H.\@  Comparison of the ATCA observations with
single-dish maps from the Mopra Telescope and sensitive spectra from
the Swedish-ESO Submillimetre Telescope indicates that the
condensation is a massive clump of $\sim$$10^4$ \Msol\ within a larger
$\sim$$10^5$ \Msol\ molecular cloud.  The clump is centered adjacent to
a compact, obscured \HII\ region which is part of a linear structure
of radio continuum sources extending across the molecular cloud.  We
suggest that the clump represents a possible site for triggered
star formation.  Examining the integrated line intensities as a
function of interferometer baseline length, we find evidence for
decreasing HCO$^+$/HCN and HCN/HNC ratios on longer baselines.  These
trends are consistent with a significant component of the HCO$^+$
emission arising in an extended clump envelope and a
lower HCN/HNC abundance ratio in dense cores.

\end{abstract}


\section{Introduction}\label{sec:intro}

Cold molecular gas in the interstellar medium (ISM) is usually studied
via rotational transitions of trace molecules that emit at millimeter
and sub-millimeter wavelengths.  The most commonly used tracer is CO,
which typically traces gas densities of
$\sim$10$^2$--10$^3$ cm$^{-3}$.  Higher densities are best probed
using molecules with larger dipole moments such as HCO$^+$ and HCN.\@
Emission from these molecules comes from the densest condensations
within molecular clouds, often referred to as ``clumps'' and ``cores''
\citep[e.g.,][]{Blitz:91}.  A molecular core has typical radius $R
\sim$ 0.1 pc, mass $M \sim$ 1--10 \Msol, and density $n \gtrsim 10^4$
cm$^{-3}$, and is thought to eventually form a single star or multiple
star system.  Clumps are somewhat larger structures which may evolve
into star clusters.  The use of high-density tracers to study the
initial conditions for star formation has long been common practice
for Galactic observers, and has recently gained greater prominence in
studies of nearby galaxies following the demonstration of a tight
correlation between HCN and far-infrared (FIR) luminosities
\citep{Gao:04b}, the latter tracing recent star formation.

Although they have similar dipole moments, HCO$^+$ and HCN may not
trace dense gas equally well, as one must consider the chemical
processes which lead to their formation and destruction.  This becomes
especially apparent when examining molecular abundances in the
Magellanic Clouds, where a lower dust-to-gas ratio and locally active
star formation activity combine to yield a stronger photoionizing flux
than in the Galaxy.  As a result, molecular clouds are immersed in
extensive photon-dominated regions (PDRs) where carbon is mostly
ionized, as evidenced by the much higher [\CII]/CO intensity ratios in
the LMC \citep{Mochizuki:94,Israel:96}.  While the formation of
HCO$^+$ in regions shielded from far-ultraviolet (FUV) radiation is
dominated by reactions between H$_3^+$ and CO, it can also occur in
PDRs via reactions between C$^+$ and OH, O$_2$, and H$_2$O
\citep{Graedel:82}.  Thus, the spatial extent of the HCO$^+$ emission
may be much larger than that of HCN---even comparable to that of CO.

The HNC molecule, on the other hand, has been suggested to be a tracer
of cold cores, as steady-state chemical models predict that it will be
much weaker than HCN in regions of high temperature and density
\citep{Schilke:92}.  Consistent with these predictions, the HCN/HNC
abundance ratio is particularly high ($\approx$80) in the hot core of
Orion KL \citep{Schilke:92}, whereas it is close to unity in cool dark
clouds \citep{Churchwell:84,Hirota:98}.  However, observations of
starburst galaxies, where temperatures are expected to be high, have
revealed HNC intensities that are sometimes close to those of HCN
\citep{Huette:95,Aalto:02}.  While an extended component of cool,
quiescent gas could contribute to the HNC emission, \citet{Aalto:02}
speculate that PDR chemistry may play a role: reactions involving
HCNH$^+$ may become important, which recombines to produce HCN and HNC
with equal probability and largely independent of kinetic temperature.
Thus, HNC is potentially another tracer of PDR conditions.

As one of the initial targets for the prototype millimeter receivers
on the Australia Telescope Compact Array\footnote{The Australia
  Telescope is funded by the Commonwealth of Australia for operation
  as a National Facility managed by CSIRO.} (ATCA), we have observed the
\HII\ region N113 in the Large Magellanic Cloud (LMC) in the 3mm
transitions of HCO$^+$, HCN, HNC, and C$_2$H.\@  The LMC's proximity
($d$=52 kpc, \citealt{Panagia:99}) and low metallicity make it a very
attractive target for studies of the molecular ISM, but its southern
declination has heretofore made it inaccesible to millimeter
interferometry.  We use HCN as a fiducial dense gas tracer and examine
the relative intensities of the other molecular lines (including
C$_2$H, produced in reactions involving C$^+$) for enhancements due to
the presence of a PDR.\@  In particular, we examine the line ratios as a
function of spatial scale, which allows a comparison of compact and
extended emission regions.  

The molecular cloud associated with N113 \citep{Henize:56} is one of
the richest in the LMC, with a peak CO(1--0) brightness temperature
(measured with the 45\arcsec\ beam of the Swedish-ESO Submillimetre
Telescope, SEST) of $\sim$9 K \citep{Chin:97}.  Also associated with
N113 are three young ($<$10 Myr) stellar clusters, NGC 1874, NGC 1876,
and NGC 1877 \citep{Bica:92}.  Current star formation activity appears
concentrated in three compact radio continuum sources, superposed on
an extended emission component and aligned in a
northwest-southeast direction with separations of $\sim$20\arcsec\
\citep{Brooks:97}.  The faintest, easternmost source contains by far
the brightest H$_2$O maser in the LMC \citep{Whiteoak:86,Lazendic:02}
as well as an OH maser \citep{Brooks:97}.  The brightest, westernmost
source also shows H$_2$O maser emission \citep{Lazendic:02}; it is
towards this source that we have targeted our ATCA observations.  All
four of the 3mm transitions observed with ATCA have been previously
detected in this region with the SEST.\@ In this paper, we present the
ATCA results and compare them with single-dish results from the Mopra
and SEST telescopes.

\section{Observations}\label{sec:obs}

\begin{figure*}
\begin{center}
\includegraphics[height=17cm,angle=-90]{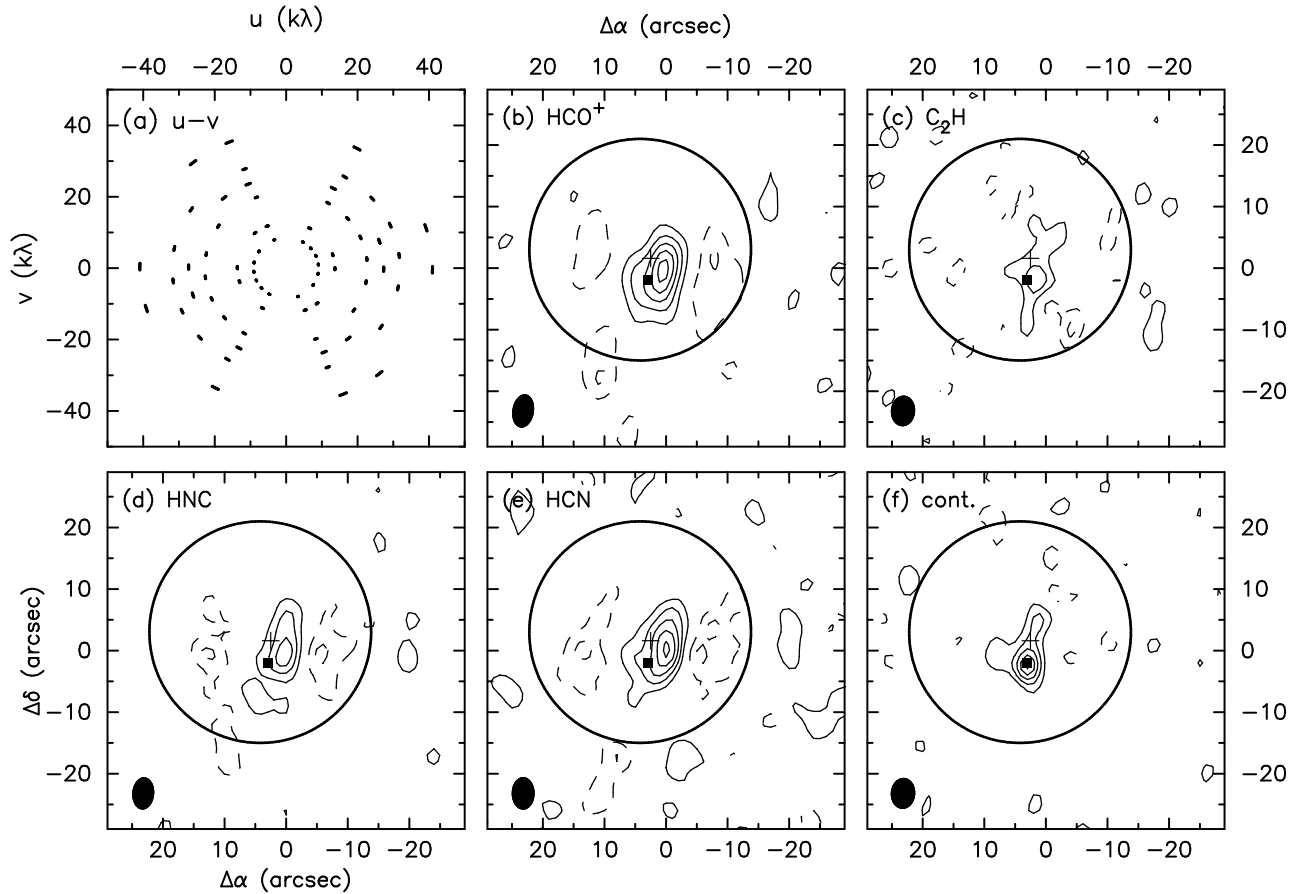}
\end{center}
\caption{
Summary of ATCA observations towards N113.  
(a) Coverage of the visibility plane for HCO$^+$ obtained from
combining both runs.
Other panels show integrated intensity maps for (b) HCO$^+$; (c) C$_2$H;
(d) HNC; (e) HCN; (f) continuum.
Offsets are relative to $\alpha_{2000}$=5$^{\rm h}$13$^{\rm m}$17\fs2,
$\delta_{2000}$=$-69$\arcdeg22\arcmin23\arcsec.   
The synthesized beam FWHM is shown in the lower left
corner and the large circle represents the FWHM field of view.  
The solid square and cross represent the positions of the $K$-band
source and the weaker H$_2$O maser respectively.
The contour levels are spaced by 6 K \kms\ for HCO$^+$ and HCN, 3 K
\kms\ for HNC and C$_2$H, and 0.06 K for the continuum.
\label{fig:mommaps}
}
\end{figure*}

\subsection{ATCA 3-mm data}

Observations with the ATCA were
made in 2003 July and August in the EW214 and EW367 configurations
respectively.  At the time of the observations, the ATCA had three
antennas of 22~m diameter equipped with dual polarisation 3-mm
receivers covering the bands 84.9--87.3 and 88.5--91.3 GHz.  At these
frequencies the primary beam has a FWHM of 36\arcsec.  Both
configurations had all three antennas arranged in an east-west line.
The EW214 configuration provided baselines of 30m, 75m, and 105m,
whereas the EW367 provided baselines of 45m, 90m, and 135m.  The
pointing center for the observations was
$\alpha_{2000}$=5$^{\rm h}$13$^{\rm m}$18$^{\rm s}$,
$\delta_{2000}$=$-69$\arcdeg22\arcmin20\arcsec.  The ATCA correlator
can be configured to observe two frequencies simultaneously, and we
cycled between 3 pairs of frequencies during each observing session in
order to obtain comparable coverage of the visibility plane.  The
first pair of frequencies consisted of the HCO$^+$ ($J$=1--0) and HNC
($J$=1--0) lines at 89.1885 and 90.6635 GHz respectively, redshifted
to the appropriate observatory frequency for $v_{\rm LSR}$=235 \kms,
the second pair consisted of the HCN ($J$=1--0) line at 88.6318 GHz
and HNC, and the third pair consisted of the C$_2$H ($N$=1--0,
$J$=3/2--1/2, $F$=2--1) line at 87.3169 GHz and a 128 MHz wide
continuum window at 86.243 GHz.  Each line was observed with a channel
spacing of 0.5 MHz across a 64 MHz bandwidth, except for C$_2$H, which
was observed with a channel spacing of 0.0625 MHz across 16 MHz.
Figure~\ref{fig:mommaps}(a) shows the coverage of the visibility plane
obtained for the HCO$^+$ line.

The data were calibrated and imaged with the MIRIAD package
\citep{Sault:95}, which at ATNF has been updated with new procedures
to handle ATCA 3-mm data.  Flux calibration was performed using
Uranus, while the quasars B0537-441 and B1253-055 (3c279) were used
for gain and bandpass calibration respectively.  However, for the last
frequency pair (C$_2$H and continuum), the SiO maser R Dor was used
for gain calibration instead; in doing so the maser flux was assumed
to be constant over the course of an $\sim$8 hour observation.  An
instrumental phase error causing phase discontinuities when changing
sources had been identified with ATCA 3-mm data,\footnote{This is
believed to be an instrumental error which corrupts the antenna
baseline solution.} and we compensated for this using an antenna
position correction which was determined by observations of multiple
quasars within a few days of the observations.  We produced channel
maps spaced in velocity by 2 \kms, which is close to the effective
resolution provided by the correlator in the HCN, HNC, and HCO$^+$
lines.  The phase center was shifted to $\alpha_{2000}$=5$^{\rm
  h}$13$^{\rm m}$17\fs2,
$\delta_{2000}$=$-69$\arcdeg22\arcmin23\arcsec\ before imaging, a
shift of 5\arcsec\ (this shift has negligible impact on the primary
beam response at the locations of detected emission).  The noise level
and synthesized beamwidth for each map are given in
Table~\ref{tbl:atcapars}.  After imaging the maps were CLEANed to a
2$\sigma$ level.

\begin{table}
\caption{Properties of the ATCA images\label{tbl:atcapars}}
\begin{tabular}{lccccccc}
Line & Beam size & Beam PA & $\sigma_{\rm ch}$\tablenotemark{a} & $\sigma_{\rm ch}$ &
$T_{\rm peak}$ & Vel.\ range & Flux\tablenotemark{b}\\
& (\arcsec\ $\times$ \arcsec) & (\arcdeg) & (mJy bm$^{-1}$) & (K) &
(K) & (\kms) & (Jy \kms) \\ \tableline
HCO$^+$ & 5.1 $\times$ 3.2 & $-9$ & 41 & 0.38 & 4.2 & [229,241] & 5.2--8.8 \\
HCN     & 5.0 $\times$ 3.4 & 0    & 42 & 0.39 & 2.8 & [223,243] & 4.1--6.9 \\
HNC     & 4.9 $\times$ 3.3 & $-5$ & 23 & 0.21 & 1.4 & [229,241] & 1.0--2.1 \\
C$_2$H  & 4.7 $\times$ 3.6 & $-5$ & 49 & 0.46 & 1.1 & [233,239] & 0.5--1.5 \\
Cont.   & 4.7 $\times$ 3.7 & $-4$ & 3 & 0.03  & 0.3 & --- & 0.04 Jy \\
\tableline
\end{tabular}
\tablenotetext{a}{RMS noise for a 2 \kms\ channel (line images) or
  continuum integrated across 84 MHz.}
\tablenotetext{b}{Flux measurements are made in boxes of
  10\arcsec\ $\times$ 10\arcsec\ and 20\arcsec\ $\times$ 20\arcsec\
  centered on $\alpha_{2000}$=5:13:17.2, $\delta_{2000}$=$-69$:22:23.}
\end{table}

\subsection{ATCA 1.3-cm data}

Additional observations were obtained with ATCA in 2005 March using
the recently installed 12-mm receivers.  N113 was observed in the H214
configuration (baseline lengths 80m--250m) over two days with a total
integration time of 160~min.  The correlator was configured for 512
channels across 8 MHz (velocity resolution 0.24 \kms) at 23.673 GHz
and 32 channels across 128 MHz at 23.9 GHz.  The narrowband
observations were designed to detect the NH$_3$(1,1) transition but
failed to do so at an RMS noise level of 20 \mjybm.  The wideband
observations, used to image the continuum, achieved a noise level of
$\sim$0.4 \mjybm.  Flux, bandpass, and phase calibration were
performed using the quasars B1934-638, B1921-293, and B0454-810
respectively, following standard reduction procedures in MIRIAD.\@
After Fourier inversion the 1.3-cm continuum image was CLEANed and
restored using a beam size of 10\farcs1$\times$9\farcs1.  Note that
because of the limited coverage of the Fourier plane achieved with the
H214 configuration, structure on scales much larger than 30\arcsec\
will be poorly imaged.  Thus, the resulting image is dominated by
compact ionized regions, and the extended continuum emission from the
large-scale \HII\ region was not detected.

\subsection{SEST data}

Although most of our comparison with SEST data makes use of values
previously presented by \citet{Chin:97}, we also made use of some data
which had not yet been published.  These consisted of small grid maps
made in the HCN ($J$=1$\rightarrow$0) and HCO$^+$
($J$=1$\rightarrow$0) lines in two observing runs with the SEST in
1997 January and July.  The maps cover a region of roughly 1\farcm5
$\times$ 1\farcm5 centered on the position $\alpha_{2000}$=5$^{\rm
h}$13$^{\rm m}$18\fs2, $\delta_{2000}$=$-69\arcdeg 22\arcmin
35\arcsec$.  The grid spacing was 20\arcsec\ (for comparison the SEST
beam has a FWHM of 58\arcsec), but the grids were not completely
sampled in the outer parts.  A 3mm SIS receiver was used in
single-sideband mode with a 2000 channel acousto-optical spectrometer
having a 43 kHz channel separation.

A dual beam switching mode, where the source is placed alternately in
the main and reference beams, was used to remove the sky background,
and the spectra were placed on a \tmb\ scale using a main beam
efficiency of 0.76.  Typical single-sideband (SSB) system temperatures
were $\sim$200~K.
The CLASS package was used to calibrate and baseline the spectra.  For
purposes of this paper we have produced high signal-to-noise spectra
for both lines by averaging the 9 spectra within $-20^{\prime\prime} <
\Delta\alpha < 20^{\prime\prime}$ and $-20^{\prime\prime} <
\Delta\delta < 20^{\prime\prime}$ relative to the center position.
The corresponding maps show only a moderately resolved central source
and are not reproduced here.  Further details of the SEST observations
will be presented in a forthcoming paper (M. Wang et al., in preparation).

\begin{figure*}
\plotone{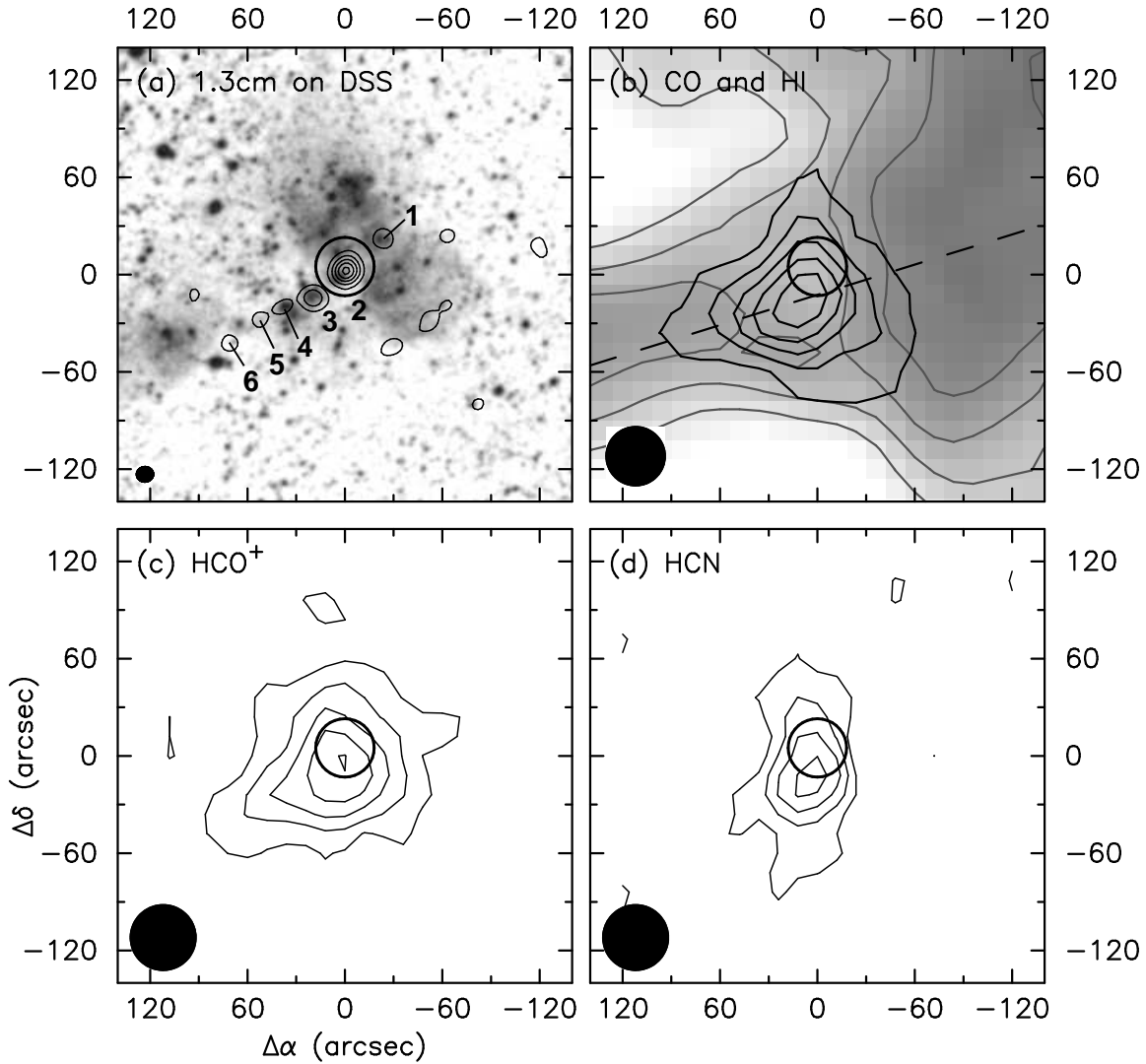}
\caption{
(a) ATCA 1.3cm map (contours) overlaid on an $R$-band image from the
  Digitized Sky Survey.  Contour levels correspond to $2n^2$
  mJy~bm$^{-1}$ ($n$=1,2,...).  
(b) Mopra CO (1--0) map (black contours) overlaid on an \HI\ peak
  intensity image (grayscale and gray contours).  Contour levels are
  10 K \kms\ for the CO map and 40, 50, ..., 80~K for the \HI\
  map. The dashed line represents the position-velocity slice shown in
  Fig.~\ref{fig:pv}.
(c) Mopra HCO$^+$ (1--0) map.  Contour levels are spaced by 0.6 K
  \kms.
(d) Mopra HCN (1--0) map.  Contour levels are spaced by 0.5 K \kms.
  In each case the solid open circle represents the ATCA pointing
  position.  Offsets are relative to $\alpha_{2000}$=5$^{\rm
  h}$13$^{\rm m}$18$^{\rm s}$,
  $\delta_{2000}$=$-69$\arcdeg22\arcmin25\arcsec.
\label{fig:mopmaps}}
\end{figure*}

\subsection{Mopra data}

A 4\arcmin\ $\times$ 4\arcmin\ region centered on the position
$\alpha_{2000}$=5$^{\rm h}$13$^{\rm m}$18$^{\rm s}$,
$\delta_{2000}$=$-69\arcdeg 22\arcmin 25\arcsec$ was mapped in the
$J$=1$\rightarrow$0 transitions of HCO$^+$, HCN, and CO with the 22-m
ATNF Mopra telescope in three runs from 2004 June to September.  The
newly implemented On-the-Fly (OTF) mode was used, in which the telescope
takes data continuously while moving across the sky.  Spectra were
spaced by 7\arcsec\ so that the 36\arcsec\ Mopra beam (33\arcsec\ at
CO) would be well oversampled in the scanning direction; the row
spacing was 10\arcsec, also oversampling the beam.  Each row was
preceded by an off-source integration at a location about 10\arcmin\
from the map centre.  Typical SSB \tsyse\ values of 200~K for HCO$^+$
and HCN and 600~K for CO were obtained in clear conditions.  The
pointing was checked on the SiO maser R Dor prior to each map; typical
corrections were less than 5\arcsec.  The digital correlator was
configured to output 1024 channels across 64 MHz in each of two
orthogonal polarisations, except for the June observations where only
one polarisation was recorded.  Initial spectral processing
(baselining and calibration onto a $T_A^*$ scale) was performed using
the {\it LiveData} task in AIPS++, and the spectra were gridded into
datacubes using the AIPS++ {\it Gridzilla} task.  During the gridding,
a Gaussian smoothing kernel with a FWHM half that of the beamsize was
convolved with the data, so the effective resolutions of the final
cubes were 40\arcsec\ for HCO$^+$ and HCN and 36\arcsec\ for CO.  The
cubes were then rescaled onto a \tmb\ scale using an ``extended beam''
efficiency of $\eta_{\rm xb}$=0.55 for CO and 0.65 for HCO$^+$ and HCN
\citep{Ladd:05}.  This takes into account that sources larger than
about 80\arcsec\ in diameter will couple to both the main beam and the
inner error beam of the telescope.

Comparison of Mopra and SEST spectra taken at the central Mopra position,
with the Mopra maps convolved to the SEST resolution, indicate that
the Mopra line intensities are lower by $\sim$30\%.  This is not
surprising given that the extended beam efficiency is higher than the
main beam efficiency $\eta_{\rm mb}$ by $\sim$30\% \citep{Ladd:05}.
While use of $\eta_{\rm mb}$ rather than $\eta_{\rm xb}$ would place
the Mopra data on a more accurate temperature scale for compact
sources, it would lead to an overestimate of the total map flux when
assuming a Gaussian beam size equal to the main beam.  Thus we
continue to use $\eta_{\rm xb}$ to calibrate the spectra, noting
however that the quoted beam size is somewhat less than the effective
beam size due to the presence of a significant inner error beam.

\section{Analysis \& Results}\label{sec:results}

\subsection{The N113 Molecular Cloud}\label{sec:mopresults}

We begin by presenting maps and derived properties for the molecular cloud as
a whole, based on the Mopra data.  We discuss the high-resolution ATCA 3-mm
data in \S\ref{sec:atcaresults}.

\subsubsection{ATCA 1.3-cm and Mopra maps}

\begin{table}
\caption{Parameters of Gaussian fits to Mopra images\label{tbl:mopfits}}
\begin{tabular}{lccccccc}
Line & Center RA & Center DEC & Major axis\tablenotemark{a} & Minor
axis\tablenotemark{a} & Position angle\\
& (J2000) & (J2000) & (\arcsec) & (\arcsec) & (\arcdeg)\\ \tableline
CO      & 5 13 19.9 & $-69$ 22 43 & 81 $\pm$ 2 & 67 $\pm$ 2 & $-80$ \\
HCO$^+$ & 5 13 19.4 & $-69$ 22 32 & 89 $\pm$ 3 & 67 $\pm$ 2 & $-59$ \\
HCN     & 5 13 19.3 & $-69$ 22 33 & 84 $\pm$ 4 & 25 $\pm$ 2 & $-4$ \\
\tableline
\end{tabular}
\tablenotetext{a}{Deconvolved using a Gaussian beam of 36\arcsec\ FWHM
  for CO and 40\arcsec\ for HCO$^+$ and HCN.}
\end{table}

Figure~\ref{fig:mopmaps}(a) shows the ATCA 1.3-cm continuum image
overlaid on a red Digitized Sky Survey image showing much of the
diffuse H$\alpha$ emission from the \HII\ region.  In addition to the
string of three bright sources reported by \citet{Brooks:97}, labelled
2, 3, and 4, we have detected three additional continuum sources along
the same formation, one west of the main source (Source 1) and two to
the southeast (Sources 5 and 6).  The fluxes of the six continuum
sources, from west to east, are estimated based on Gaussian fits to be
6.2, 66, 18, 4.1, 3.9, and 4.4 mJy.  Measuring the flux by direct
summation of image pixels yields fluxes consistent with these values
within 1 mJy.  Apparent sources which lie outside this linear
structure (e.g., southwest of the image center) are likely to be
sidelobes that were incompletely removed.

Figures~\ref{fig:mopmaps}(b)--(d) show the Mopra CO, HCO$^+$, and HCN
maps integrated from a velocity of 231 to 240 \kms.  Remarkably, at
the resolution of Mopra the molecular cloud appears to be a centrally
condensed structure, in contrast to the filamentary structure which is
characteristic of the \HI\ clouds in the LMC \citep{Kim:03}.  The CO
map is shown overlaid on \HI\ emission contours from the combined
Parkes/ATCA survey \citep{Kim:03}.  The HCO$^+$ emission imaged with
Mopra shows a similar extent to the CO, however the peak intensity is
shifted $\sim$10\arcsec\ to the north.  The HCN emission appears to be
narrower in the east-west direction than the other tracers, as
confirmed by fitting of two-dimensional Gaussians to the images
(Table~\ref{tbl:mopfits}).

\subsubsection{Mass and Dynamical Properties}\label{sec:mass}

We can estimate the mass of the molecular cloud from the CO data in
two ways.
Adopting a conversion from K to Jy of 1 K = $9.6 \times 10^{-3}$ Jy
arcsec$^{-2}$, the total CO(1--0) line flux detected by the Mopra
observations is $2.9 \times 10^3$ Jy \kms\ over a 4\arcmin\ $\times$
4\arcmin\ region.  Using a standard (Galactic) CO to H$_2$ conversion
factor of $N_{\rm H_2}/I_{\rm CO} = 2.0 \times 10^{20}$ cm$^{-2}$ (K
\kms)$^{-1}$ \citep{Strong:96,Dame:01}, this translates into a total
gas mass (including He) of $8.2 \times 10^4$ \Msol.  We can compare
this with a virial mass estimate, following \citet{Heikkila:99}, of
$M_{\rm vir} = 151\, R_{\rm CO}(\Delta v)^2$, where $R_{\rm CO}$ is the
deconvolved FWHM size of the cloud in pc.  For N113, a Gaussian fit to the
integrated CO intensity gives $R = 74\arcsec$ (19 pc) after
deconvolution, which for $\Delta v$ = 5.2 \kms\ (the average FWHM
linewidth over the inner 1\arcmin\ $\times$ 1\arcmin) yields $M_{\rm
vir} = 7.8 \times 10^4$ \Msol.  Although only a rough estimate, this
suggests, as noted previously by \citet{Chin:97}, that a standard
conversion factor yields mass estimates for LMC molecular clouds that
are consistent with virial mass estimates when observed at resolutions
of $\sim$10 pc.  A similar result was obtained for SMC clouds by
\citet{Rubio:93}.

Using the mass estimate above and a radius of 15~pc (estimated as 80\%
of the FWHM, following \citealt{Heikkila:99}) gives an overall density
for the cloud of $\left<n_{\rm H}\right> \sim 200$ cm$^{-3}$.  This is
significantly less than the critical density for excitation of HCO$^+$
and HCN ($\sim$$10^5$ cm$^{-3}$), indicating that the material in the
cloud is strongly clumped.  Of course, our ATCA observations
(described below) provide direct evidence of this clumping.

\begin{figure}
\includegraphics[height=8.5cm,angle=-90]{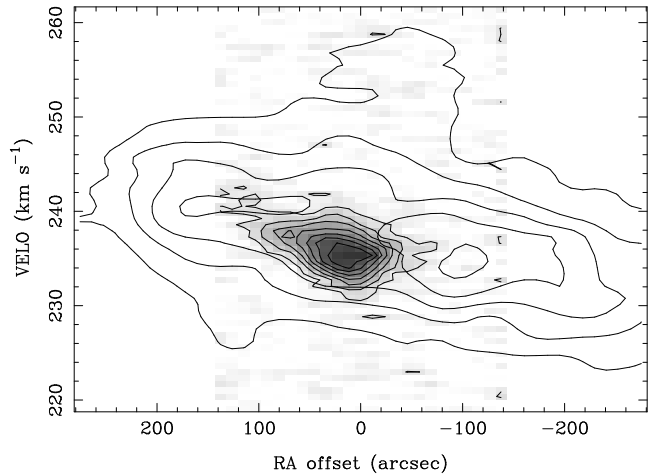}
\caption{
Position-velocity slice through the Mopra CO data cube (grayscale with
light contours) with \HI\ contours overlaid.  The \HI\ data have a
resolution of 60\arcsec\ FWHM.  The slice is centered on
$\alpha_{2000}$=5:13:18, $\delta_{2000}$=$-69$:22:43 and has a position
angle of 107\arcdeg.
Contour levels are spaced by 0.5~K for CO and 15~K for \HI.
\label{fig:pv}}
\end{figure}

Inspection of the CO datacube reveals a velocity gradient of $\sim$5
\kms\ from the center to the southeast extension.  The presence of
such a gradient might be interpreted in terms of rotation
\citep[e.g.,][]{Rosolowsky:03}.  However, the CO position-velocity
diagram (Figure~\ref{fig:pv}) is not strongly indicative of rotation,
but rather suggests a high velocity dispersion core that blends into a
low velocity dispersion wing with a slightly different velocity
centroid.  We note that the wing emission could be composed of
unresolved, clumpy structure.  An overlapping slice through the \HI\
datacube of \citet{Kim:03}, smoothed to 1\arcmin\ resolution, shows a
similar velocity gradient across a filament-like structure 10\arcmin\
(150 pc) in length.  We conclude that the CO velocity gradient is not
due to the rotation of a kinematically decoupled structure, but
instead reflects the bulk motion of the gas from which the molecular
cloud formed. 

The CO cloud appears to sit within a local minimum of \HI\ peak brightness
temperature [Fig.~\ref{fig:mopmaps}(b)].  This does not necessarily
require a a large-scale conversion of the \HI\ into H$_2$, as the \HI\
at the location of the cloud still has a high integrated intensity.
Rather, the lower peak $T_b$ is mainly the result of a flat-topped
line profile, which may indicate absorption by colder, optically thick
\HI.

\begin{table*}
\caption{Comparison of Mopra and ATCA fluxes\label{tbl:fluxes}}
\begin{center}
\begin{tabular}{lcccccc}
Line & Mopra (tot) & Mopra (pb) & ATCA (im) & ATCA (uv) & $a_{\rm maj}$ (uv) 
& $a_{\rm min}$ (uv) \\ \tableline
CO & 2900 & 320 & --- & --- & --- & --- \\
HCO$^+$ & 150 & 17 & 8.8$\pm$0.2 & 14.2$\pm$0.9 & 7\farcs7$\pm$0\farcs8 &
5\farcs4$\pm$0\farcs8 \\ 
HCN & 73 & 10 & 6.9$\pm$0.2 & 10.5$\pm$0.9 & 6\farcs3$\pm$2\farcs2 &
4\farcs8$\pm$2\farcs2 \\ 
HNC & --- & --- & 2.1$\pm$0.1 & \phn3.9$\pm$0.5 & 6\farcs8$\pm$3\farcs0 &
5\farcs2$\pm$2\farcs4\\ 
C$_2$H & --- & --- & 1.5$\pm$0.2 & \phn3.9$\pm$1.1 & 7\farcs0$\pm$1\farcs7
& --- \\
Cont. & --- & --- & 42 mJy & 47$\pm$7 mJy & 2\farcs6$\pm$0\farcs6 &
--- \\
\tableline
\end{tabular}
\tablecomments{All fluxes except continuum are in Jy \kms.  Mopra
  fluxes are derived from the total image (2nd col.) and restricted to
  the ATCA field of view (3rd col.).  ATCA fluxes are derived from the
  central 10\arcsec\ $\times$ 10\arcsec\ of the image and from a
  Gaussian fit to the visibility data.  The Gaussian has been
  restricted to be circular for C$_2$H and the continuum.}
\end{center}
\end{table*}

\subsection{The Dense Embedded Clump}\label{sec:atcaresults}

\subsubsection{ATCA 3-mm maps}

Figures~\ref{fig:mommaps}(b)--(f) show the integrated intensity maps
for the four lines observed with ATCA as well as the continuum at 86
GHz.  Both the line and continuum emission appear to be concentrated
in a single emission peak, or ``clump.''  These maps were produced by
directly summing the velocity channels over the ranges given in
Table~\ref{tbl:atcapars}.  Due to the presence of large negative
sidelobes in the maps, likely resulting from gain calibration errors
and imperfect deconvolution, the integrated fluxes in
Table~\ref{tbl:atcapars} are given for both the inner 10\arcsec\
$\times$ 10\arcsec\ and 20\arcsec\ $\times$ 20\arcsec\ (larger values
correspond to the smaller box), and should be considered indicative
only.  Also marked on each panel of Figure~\ref{fig:mommaps} are the
positions of the obscured 2MASS point source 05131777-6922250 and the
H$_2$O maser detected by \citet{Lazendic:02}.  Note that the infrared
source coincides with the continuum source but is offset from the peak
of the molecular line emission.

\subsubsection{Flux Recovered by ATCA}

Table~\ref{tbl:fluxes} lists the total fluxes detected in the Mopra
and ATCA maps, as well as the flux detected by Mopra with the ATCA
primary beam response applied (i.e., the Mopra flux restricted to the
ATCA field of view).  We caution that the Mopra flux within a single
primary beam may be overestimated if emission outside the beam is smoothed
into it; it may also be underestimated because of the use of
$\eta_{\rm xb}$ to calibrate the data.  In any case, we find that the
ATCA flux within the central 10\arcsec\ $\times$ 10\arcsec\ of the
integrated HCO$^+$ and HCN maps is 50--70\% of the Mopra flux within
the ATCA primary beam.  For an independent comparison, we can also
extrapolate the ATCA flux to short (``single-dish'') baselines by
fitting a two-dimensional Gaussian source in the visibility domain.
As shown in the last three columns of Table~\ref{tbl:fluxes}, which
give the fitted flux and FWHM major and minor axes, this
yields a total flux that is consistent with the single-dish flux
within the ATCA primary beam (assuming $\sim$20\% calibration uncertainties
for both telescopes).  The difference compared to measuring the flux
in the image plane is likely due to a more inaccurate
extrapolation to short baselines provided by CLEAN, coupled with
uncorrected phase errors that lead to negative regions in the
deconvolved images.

The analysis above suggests that the more extended component of
HCO$^+$ or HCN emission which does not appear in the ATCA images can
be inferred by inward extrapolation of the visibility amplitude.
Assuming all of the flux is recovered by the visibility plane fitting,
we derive an approximate beam filling fraction (within the 36\arcsec\
ATCA beam) of 3\% for the HCO$^+$ and HCN emission.  Of course, this
applies only to the region observed with the ATCA, which contains just
$\sim$12\% of the total flux detected by Mopra.  Clearly it would be
of great interest to cover the entire molecular cloud with an ATCA
mosaic, to determine what fraction of the total line emission the
interferometer would be sensitive to.

\begin{figure}
\includegraphics[width=8cm]{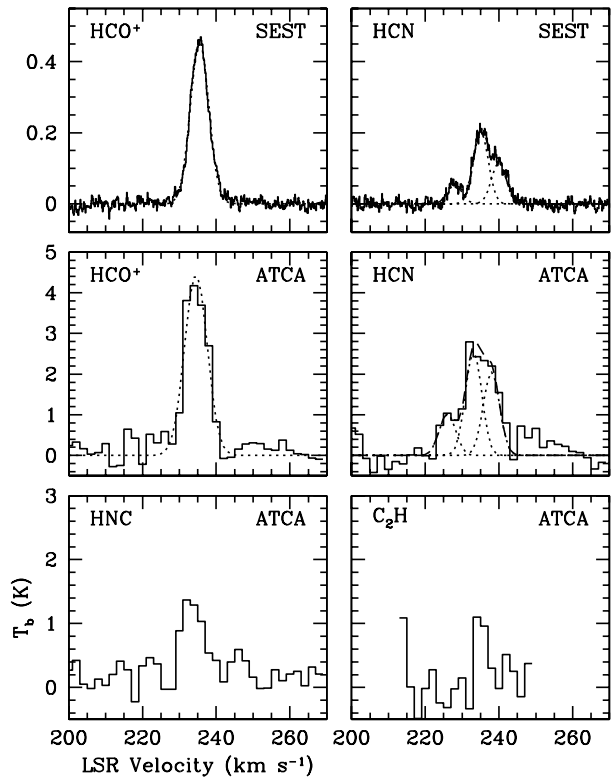}
\caption{
Comparison of ATCA and SEST spectra.  SEST spectra, in $T_{\rm mb}$
units (top two panels), have been averaged over a 40\arcsec\ $\times$
40\arcsec\ region centred on $\alpha_{2000}$=5:13:18,
$\delta_{2000}$=$-69$:22:35.  Spectra through the ATCA datacubes,
smoothed to 2 \kms\ and converted to brightness temperature units
using the synthesized beam size, have been taken at the nominal peak
position of $\alpha_{2000}$=5:13:17.2, $\delta_{2000}$=$-69$:22:23.
Gaussian fits to the HCO$^+$ and HCN lines (the latter taking into
account the spacing of the three hyperfine components) are shown as
dotted lines.
\label{fig:atspec}}
\end{figure}

\begin{figure}
\includegraphics[width=8.5cm]{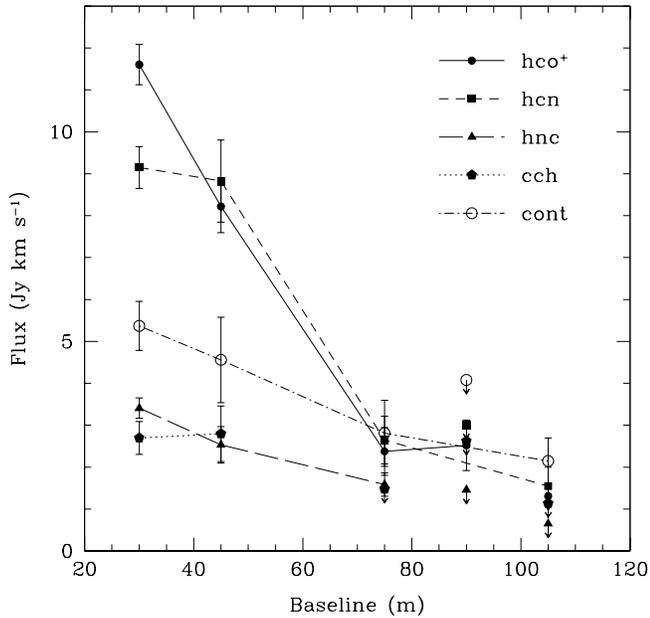}
\caption{
Integrated line flux for HCO$^+$, HCN, HNC, and C$_2$H as a
function of projected ATCA baseline.  Visibilities have been averaged
within bins of 15m width.  For the continuum, fluxes are in units of
10 mJy.  Plot symbols with arrows indicate 3$\sigma$ upper limits and
are not connected with lines.
\label{fig:uvamp}}
\end{figure}

\subsubsection{Fitting of ATCA and SEST spectra}\label{sec:specfit}

While the Mopra OTF maps of HCN and HCO$^+$ cover a larger area, their
sensitivity at any given position is a factor of 3--4 worse than the
pointed SEST spectra, so we have opted to derive line widths and hyperfine
ratios from the SEST data for comparison with the ATCA data.

The averaged SEST spectra are shown in Figure~\ref{fig:atspec}, along
with ATCA spectra taken through the central 1\arcsec\ pixel of each datacube.
The HCO$^+$ and HCN brightness temperatures $T_b$ measured in the ATCA
maps are a factor of $\sim$10 higher than the SEST measurements, which
is not surprising given the presence of clumping and the factor of
$\sim$100 difference in beam area.  We have also made single-component
Gaussian fits to the HCO$^+$ spectra and three-component fits to the
HCN spectra; the latter assume that the two satellite hyperfine
transitions ($F$=0--1 and $F$=1--1) are separated by $-7.07$ and 4.85
\kms\ from the main component and have the same dispersion as the main
component.  We find for HCO$^+$ a FWHM linewidth (corrected for
instrumental broadening) of 5.8 \kms\ for SEST and 5.5 \kms\ for ATCA.
For HCN the linewidths are 4.6 \kms\ and 5.2 \kms\ for SEST and ATCA
respectively, though these are much more uncertain, being based on a
multi-component fit.  We also find a velocity offset of $-2.0$ \kms\
in the ATCA HCN spectrum relative to SEST, which we attribute to
limited spectral resolution and signal-to-noise in the ATCA data.

The ratios between the hyperfine components of HCN can be used as a
diagnostic of the optical depth, if LTE excitation is assumed
\citep[e.g., Fig.~2 in][]{Harju:89}.  In the optically thin case, the
ratio of the $F$=0--1/$F$=2--1 transitions, referred to as $R_{02}$,
should be 0.2, whereas the ratio of the $F$=1--1/$F$=2--1 transitions,
referred to as $R_{12}$, should be 0.6.  Fitting the SEST spectrum
yields $R_{02}$=0.26$\pm$.01 and $R_{12}$=0.57$\pm$.03, where the
errors are only the formal errors on the fit and thus a lower
estimate.  These appear consistent with optically thin emission.  On
the other hand, for the ATCA spectrum we find $R_{02}$=0.40$\pm$.12
and $R_{12}$=0.79$\pm$.19, indicating higher opacities for the dense
clump at the 1$\sigma$--2$\sigma$ significance level (note however
that {\it both} ratios are greater than their LTE values).  While
still requiring confirmation with higher spectral resolution data
and/or observations of rare isotopomers, this suggests the presence of
an optically thinner HCN component which dominates the total flux.  We
suggest that, rather than constituting a truly diffuse interclump
medium, this component is an enshrouding ``clump envelope'' that
accounts for the difference between the fluxes measured in the ATCA
images and those inferred by visibility fitting
(Table~\ref{tbl:fluxes}).

\subsubsection{Clump size and line width}\label{sec:clumprop}

The visibility fits given in Table~\ref{tbl:fluxes} indicate that the
HCO$^+$ emission detected by the ATCA comes from a region 8\arcsec\
$\times$ 5\arcsec\ (2.0 $\times$ 1.3 pc) in size (FWHM), with the HCN
emission coming from a somewhat smaller region.  This implies a radius
of $\sim$1.6 pc, comparable to that of the largest clumps which are
found in Galactic massive star-forming regions
\citep{Mookerjea:04,Reid:05}, and much larger than the typical size
($\sim$0.1 pc) of a core (a condensation which forms only a single
star or a multiple star system).  We therefore believe it is
appropriate to refer to this object as a ``clump.''  We note, however,
that it is common in the literature \citep[e.g.,][]{Plume:97} to use
the term ``massive cores'' to refer to the birthplaces of the young
stellar clusters where massive stars typically reside.

It is difficult to estimate the mass of the clump from the observed
line fluxes given considerable uncertainties in the excitation,
abundances, and optical depths of HCO$^+$ and HCN.  However, we can
derive a virial mass as in \S\ref{sec:mass}, this time using $R$=1.6
pc and $\Delta v$=5.5 \kms: $M_{\rm vir} = 151\, R (\Delta v)^2 = 7.3
\times 10^3$ \Msol, or about 10\% of the total cloud mass inferred
from the CO data.  Alternatively, \citet{Gao:04a} argue on the basis
of large velocity gradient (LVG) calculations that the mass of dense
gas traced by HCN satisfies
\[M_{\rm dense}({\rm H}_2) \lesssim 25 \left(\frac{L_{\rm
    HCN}}{\rm K\;km\;s^{-1}\;pc^{-2}}\right) M_\odot\;.\]
They obtain this result assuming an HCN abundance of $2 \times
10^{-8}$, a kinetic temperature of 20--50~K and an intrinsic HCN
brightness temperature of $T_b > 5$~K.  Using this formula gives an
upper limit of $1.3 \times 10^4$ \Msol\ for the N113 clump, within a
factor of 2 of the virial estimate.  The agreement between the two
mass estimates, along with the high densities implied by the detection
of HCN, HCO$^+$, and HNC, suggests that the assumption of virialization
is reasonable.  The average density of the clump is
inferred to be $\left<n_{\rm H}\right> \approx 10^5$ cm$^{-3}$, close to the
critical densities for HCN and HCO$^+$.


\begin{table}
\caption{Integrated line intensity ratios towards the N113 peak
\label{tbl:ratios}}
\begin{tabular}{lcccc}
Ratio & SEST\tablenotemark{a} & ATCA 30m & ATCA 45m & ATCA 75m\\ \tableline
HCO$^+$/HCN & 1.35$\pm$0.06 & 1.27$\pm$0.09 & 0.93$\pm$0.13 &
0.90$\pm$0.29\\
HCN/HNC & 2.81$\pm$0.21 & 2.68$\pm$0.24 & 3.49$\pm$0.71 &
1.67$\pm$0.47\\
C$_2$H/HCN & 0.29$\pm$0.02 & 0.30$\pm$0.05 & 0.32$\pm$0.08 & $<$0.56 \\
\tableline
\end{tabular}
\tablenotetext{a}{From \citet{Chin:97}.}
\end{table}

\subsubsection{Line ratio analysis}

Figure~\ref{fig:uvamp} shows the integrated flux (in Jy \kms) for all
four lines observed with ATCA as a function of the projected baseline.
The equivalent curve for the continuum flux (in Jy, scaled down by
100) is also shown.  Since the ATCA spacings are roughly in multiples
of 15m, data have been averaged in bins of that width.  As expected
for a resolved source, the fluxes decrease on longer baselines, except
for C$_2$H, which shows roughly equal fluxes on 30m and 45m baselines
(it is not detected on longer baselines, however).  The relatively
faster decline seen for HCO$^+$ compared to HCN is responsible for the
larger source size inferred from the visibility fitting
(Table~\ref{tbl:fluxes}) and is corroborated by the Mopra maps.  For
the shorter ATCA baselines, we have tabulated several integrated line
ratios, $I$(HCO$^+$)/$I$(HCN), $I$(HCN)/$I$(HNC), and
$I$(C$_2$H)/$I$(HCN), in Table~\ref{tbl:ratios}.  These are given
alongside the SEST line ratios reported by \citet{Chin:97}.  Note that
for C$_2$H we only include the 87.3169 GHz line in these calculations.

\section{Discussion}\label{sec:disc}

\subsection{Comparison with Galactic Clumps}

Recently a number of papers have analyzed the sub-structure of
molecular clouds hosting massive star formation in our Galaxy
\citep[e.g.,][]{Shirley:03,Faundez:04}.  The clumps found in these
studies have typical masses of 100--1000 \Msol, sizes of 0.25--0.5
pc, and peak densities up to $10^6$ cm$^{-3}$.  The clump we have
imaged in N113, with a radius $R \sim 1.6$ pc and mass $M \sim 10^4$
\Msol, lies at the upper end of this distribution in terms of size and
mass.  We believe our size estimate to be realistic: while the
resolution of the ATCA maps ($\sim$4\arcsec) is only slightly smaller
than the deconvolved size of the clump (6\arcsec), the visibility
amplitude as a function of baseline indicates that the source is
well-resolved (Fig.~\ref{fig:uvamp}).  It is tempting to speculate
that such massive clumps may be responsible for the formation of
``populous'' clusters of $10^3$--$10^4$ stars in the LMC
\citep{vdB:91}.

Consistent with some previous studies of massive star-forming regions
\citep[e.g.][]{Plume:97}, we do not find evidence of a size-linewidth
relation of the form $\Delta v \propto R^{0.5}$ as seen in low-mass
star-forming regions \citep{Larson:81}.  Indeed, the velocity
dispersion of the clump is comparable to or even larger than that of the
molecular cloud as a whole ($\sim$5 \kms).  This may indicate a high
level of turbulence, as found in massive star-forming cores in the
Galaxy.  However, we caution that the observed line width could be
affected by optical depth effects, and confirmation of this result
with an optically thin tracer is required.

\subsection{The Star-forming Enviroment in N113}

The overall picture that emerges of the interstellar medium near N113
is of an environment which has been strongly affected by recent star
formation.  In addition to the diffuse H$\alpha$ seen in optical
images, the linear sequence of 1-cm continuum sources, including two
which are associated with H$_2$O maser emission, point to the presence
of young ionizing stars [Fig.~\ref{fig:mopmaps}(a)].  The brightest
H$_2$O maser is found near a relatively weak ($\sim$4 mJy) continuum
source (No.\ 4) southeast of the ATCA field, where there is still substantial
HCO$^+$ and CO emission but no appreciable HCN.  Moreover, referring
to the 2MASS Point Source Catalog \citep{Skrutskie:06}, the
associated near-infrared source is apparent in all three 2MASS bands
($m_J = 13.87$, $m_H = 13.46$, $m_K = 12.71$), suggesting it is
relatively unobscured.  These two lines of evidence point to an object
that is somewhat more evolved than the brightest ($\sim$66 mJy) of the
continuum sources, which is near the peak of the molecular line
emission and detected in $K$-band but not $J$-band by 2MASS.  This
obscured source ($m_J > 14.97$, $m_H = 14.92$, $m_K = 13.25$) is shown
by a solid square in Fig.~\ref{fig:mommaps}.  A third line of evidence
is the presence of an OH maser in the former object, again suggesting
a later stage of evolution.

Assuming the brightest continuum source emits optically thin free-free
emission (spectral index $-0.1$), the measured fluxes at 24 and 86 GHz
(66 and 42 mJy respectively) correspond to an ionizing photon output
of 1.5--3 $\times 10^{49}$ s$^{-1}$ \citep{Condon:92}, equivalent to
one or two O6 stars \citep[e.g.,][]{Turner:98}.  Although the measured
fluxes suggest a spectral index of $-0.3$ instead of $-0.1$, the larger
calibration uncertainties of the 86 GHz data (a result of higher
\tsyse\ and relying on an SiO maser for gain calibration), possibly
combined with resolving out some extended flux, could explain this
discrepancy.  The 86 GHz image also appears to show an elongation to the
north, whose significance is unclear but is interesting in light
of the presence of an H$_2$O maser $\sim$4\arcsec\ north of the
$K$-band source.  The maser and associated continuum emission could
signify an outflow from the central ionizing source, but given their
large projected distance from that source ($\sim$1 pc), are more likely
to trace other ionizing stars in the same association.

Regardless of the exact nature of the radio continuum emission, its
location only 3\farcs6 arcsec (0.9 pc) from the dense molecular clump
will surely affect the lifetime of the clump.  One possibility is that
the pressure of the expanding \HII\ region may compress the clump and
thereby incite gravitational collapse.  \citet{Whitworth:94} discuss
how shocked gas layers generated by expanding \HII\ regions tend to
fragment into $\sim$7 \Msol\ condensations which can become
gravitationally unstable, yielding a self-propagating mode of star
formation.  Such a scenario might explain the alignment of
star-forming regions seen in the continuum.  On the other hand,
\citet{Gorti:02} discuss how FUV radiation surrounding an \HII\ region
heats the clumpy molecular gas and eventually photoevaporates it,
limiting the efficiency of star formation.  A clump as massive as that
inferred by our ATCA observations would not be easily photoevaporated,
but might initially experience a shock compression due to heating and
expansion of its outer layers.  Observations of shock tracers in this
region with {\it Spitzer} may be able to directly trace this process
and assess its ability to trigger new star formation.

\begin{figure*}
\includegraphics[width=17cm]{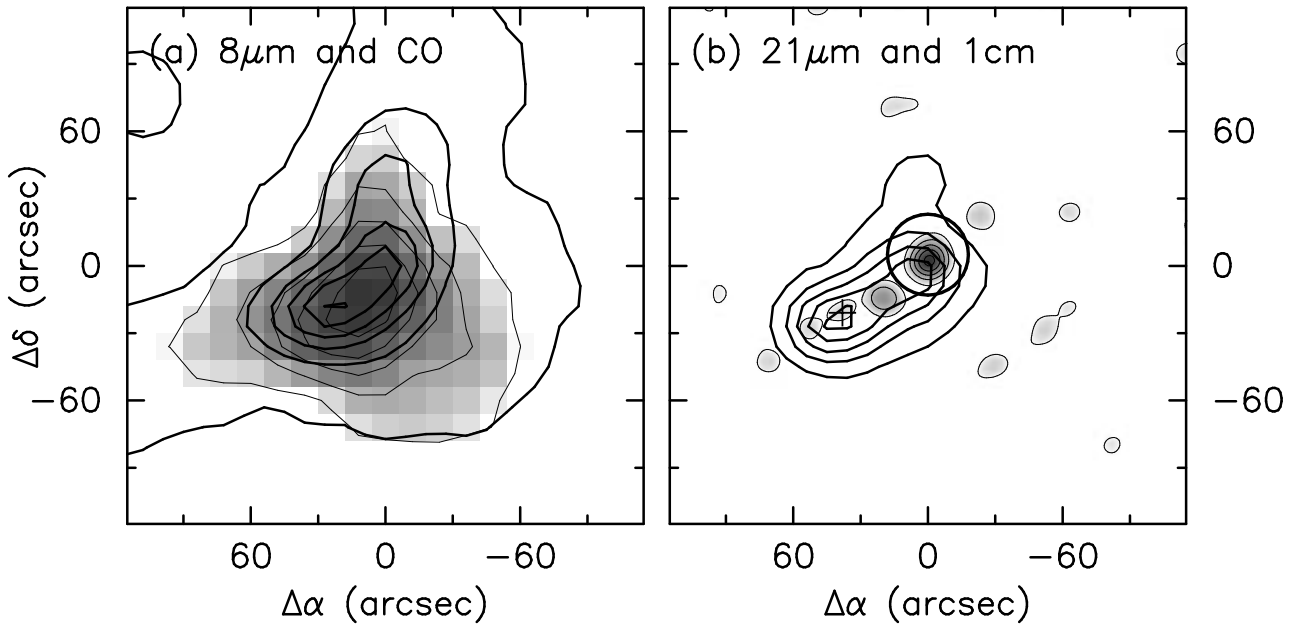}
\caption{
(a) 8$\mu$m MSX image (heavy contours) overlaid on the Mopra CO map
  from Fig.~\ref{fig:mopmaps} (light contours and grayscale).
  Contour levels for the MSX image are spaced by $1.5 \times 10^{-10}$
  W cm$^{-2}$ sr$^{-1}$.
(b) 21$\mu$m MSX image (heavy contours) overlaid
  on the 1cm continuum map (light contours and grayscale).
  Contour levels for MSX image are spaced by $4 \times 10^{-10}$ W
  cm$^{-2}$ sr$^{-1}$, while the continuum image has contours of
  $2n^2$ mJy~bm$^{-1}$ ($n$=1,2,...).  The ATCA pointing position is
  shown as a solid circle.
  The position of the bright H$_2$O maser is shown by a cross.
\label{fig:msx}}
\end{figure*}

\subsection{Evidence for PDR Chemistry}

The large HCO$^+$/HCN ratio observed with the shortest ATCA baseline
(and in the Mopra data), decreasing on longer interferometer
baselines, is consistent with the HCO$^+$ emission originating from a
more extended region than HCN.  Similar conclusions for the LMC were
reached by \citet{Johansson:94}, based on SEST mapping of the N159W
cloud, and \citet{Heikkila:99}, based on small maps of several
additional clouds.  As pointed out by \citet{Heikkila:99}, two of the
major chemical pathways for the formation of HCO$^+$ involve C$^+$,
which is expected to be abundant near the surfaces of clouds exposed
to FUV radiation, i.e. in photon-dominated regions (PDRs).  Thus,
given the higher [\CII]/CO flux ratios in the LMC \citep{Mochizuki:94},
and especially towards bright \HII\ regions \citep{Israel:96}, HCO$^+$
emission likely traces lower density gas in the LMC in addition to
well-shielded dense cores.  The situation is less clear for HCN: while
it can be produced by endothermic reactions between CN and H$_2$ in
the outer layers of a PDR, it is easily be dissociated back to form CN
in the presence of strong FUV radiation \citep{Sternberg:95}.  We
discussed above (\S\ref{sec:specfit}) the possibility that the more
optically thin HCN arises from a ``clump envelope'' which is not fully
imaged in our ATCA data.  The larger size of the HCO$^+$ emission
region then implies a somewhat larger envelope in this line than in
HCN.

C$_2$H is also expected to be enhanced when C$^+$ is, and is
especially strong relative to CO in the actively star-forming cloud
N159W \citep{Heikkila:99}.  However, the SEST and ATCA observations
show a roughly constant $I$(C$_2$H)/$I$(HCN) from large to small
scales, suggesting that C$_2$H may trace dense gas rather than the PDR
region.  Such a result is rather surprising, given the relatively low
dipole moment of C$_2$H, which should lead to its excitation in even
low-density regions.  An alternative explanation is that C$_2$H
appears less extended than HCO$^+$ because of lower optical depth.
The hyperfine structure of C$_2$H ($N$=1--0, $J$=3/2--1/2) consists of
two lines ($F$=2--1 and $F$=1--0) separated by 12 MHz.  The integrated
intensity ratio of the 2--1 to 1--0 line measured by SEST is 1.8
\citep{Chin:97}, close to the theoretical value of 2
\citep{Tucker:74}, and indicating that the opacity is relatively low.
On the other hand, the HCO$^+$/H$^{13}$CO$^+$ intensity ratio towards
N113 as measured by SEST is 34 \citep{Chin:96}, consistent with the
isotopic $^{12}$C/$^{13}$C ratio derived by \citet{Johansson:94} for
N159.  Thus, the optical depth in HCO$^+$ is likely to be small as
well.  More sensitive observations will be needed to draw firmer
conclusions on how C$_2$H is distributed relative to HCN and HCO$^+$.

A more direct indicator of PDR conditions may be the infrared emission
at 8$\mu$m, which is primarily attributed to large organic molecules
known as polycyclic aromatic hydrocarbons (PAHs) heated by far UV
radiation in the outer layers of the cloud.  As
Figure~\ref{fig:msx}(a) indicates, the distribution of
6.8$\mu$m--10.8$\mu$m emission near N113, as measured by the MSX
satellite \citep{Price:98}, closely matches the CO distribution in
extent.  This provides further evidence that the extended HCO$^+$
emission, which follows the CO distribution, is enhanced due to PDR
chemistry.  The 21$\mu$m emission, which traces heated dust, is peaked
slightly to the southeast of the PAH emission, and corresponds well to
the 1.3-cm radio continuum emission [Figure~\ref{fig:msx}(b)].

\subsection{Comparison of HCN and HNC}

Aside from a highly uncertain measurement for the 45m baseline, the
HCN/HNC intensity ratio appears to decrease on longer baselines, which
sample smaller scale structure.  This would a confirm a prediction
by \citet{Chin:97}, that the ratio would be large in warm gas subject
to strong UV heating but approach unity in cloud cores.  The
difference may be due to additional neutral-neutral pathways for HCN
formation that occur in a partially ionized medium, such as CH$_2$ + N
$\rightarrow$ HCN + H \citep{Goldsmith:81}.  These produce HCN faster
than the equivalent reactions (NH$_n$ + C $\rightarrow$ HNC +
H$_{n-1}$) produce HNC, due to the higher proportion of CH$_n$
radicals and the tendency for C to be ionized in PDRs.  However, the
ability of such reactions to produce HCN/HNC abundance ratios much
different from unity has been questioned \citep{Herbst:00}.  In any
case, the observed ratios are consistent with the HCN/HNC ratios of
$\sim$1--6 observed in starburst galaxies \citep{Aalto:02}.  A major
uncertainty in interpreting intensity ratios as abundance ratios
is the assumption of low optical depth.  The decreasing HCN/HNC on
small scales coincides with an apparent increase in the HCN opacity,
which could affect the line profiles through saturation or
self-absorption.

\section{Summary \& Conclusions}\label{sec:conc}

We have presented high-resolution imaging of HCO$^+$, HCN, HNC, and
C$_2$H towards the N113 \HII\ region in the LMC, performed with the
ATCA interferometer, as well as single-dish mapping of the CO,
HCO$^+$, and HCN lines with Mopra.  Our main conclusions are summarized
below.

\begin{enumerate}

\item On scales of several arcminutes, the molecular cloud ($M \sim
  10^5$ \Msol) shows a weak velocity gradient which is
  shared with an \HI\ filament which encompasses it.  Several radio
  continuum sources lie along the center of the filament, making it
  the locus of recent star formation.

\item Near the brightest continuum source, which we interpret as a
  compact \HII\ region, we detect a massive clump of dense molecular
  gas whose peak is slightly offset from the 3mm continuum peak.  The
  clump has a radius of $\sim$1 pc, and if in virial equilibrium, a
  mass of $\sim$10$^4$ \Msol, about 10\% of the cloud's mass.  This
  would imply a density of $\sim$10$^5$ cm$^{-3}$.  With the ionizing
  flux of 1--2 O6 stars, the compact \HII\ region may possibly trigger
  the formation of new stars in the clump.

\item While the clump---when combined with a more diffuse envelope
  inferred from visibility fitting---may account for most of
  the HCO$^+$ and HCN emission within the ATCA primary beam, the
  existence of additional clumps within the cloud is not excluded
  given the small extent of the region mapped thus far with the ATCA.

\item A decreasing HCO$^+$/HCN ratio on longer interferometer
  baselines, which probe smaller scales, suggests that the HCO$^+$
  emanates in part from a more diffuse region then the HCN.
  This is consistent with the larger
  spatial extent of the HCO$^+$ emission in the Mopra map and its
  agreement with the CO and 8$\mu$m morphology.

\item We also see evidence for a decrease in the HCN/HNC ratio on
  smaller scales.  This may reflect a lower ratio in cold, dense
  regions, as found in the Galaxy, but the interpretation may be
  clouded by an increased opacity in the HCN line.

\end{enumerate}

These observations demonstrate that maps of LMC molecular clouds on
scales of a few pc can be used to investigate the internal structure
and chemistry of clouds for comparison with Galactic samples.  With
the ATCA now covering a frequency range of 85--105 GHz with five
antennas, further observations will soon achieve better sensitivity
for a larger number of molecular lines, allowing for a more detailed
comparison with PDR models.  Ultimately, the ability to
observe both low and high excitation lines at high resolution with the
Atacama Large Millimeter Array (ALMA) will facilitate a comprehensive
understanding of the properties of molecular clouds in an external
galaxy.

\acknowledgements

We thank L. Staveley-Smith and S. Kim for providing the \HI\ datacube
used for comparison, and M. Burton and C. Henkel for providing helpful
comments on the manuscript.  Paul Ho and an anonymous referee also
suggested improvements for the revised version.  We thank R. Sault and
the staff of the Narrabri observatory for assistance and
troubleshooting at both ATCA and Mopra.  The Mopra data were obtained
in the UNSW time allocation with the kind permission of M. Burton.  TW
acknowledges support from an ARC-CSIRO Linkage Postdoctoral
Fellowship.  YNC acknowledges financial support through National
Science Council of Taiwan grant NSC 89-2119-M-032-001.  
This publication makes use of data products from the Two Micron All
Sky Survey, which is a joint project of the University of
Massachusetts and the Infrared Processing and Analysis
Center/California Institute of Technology, funded by the National
Aeronautics and Space Administration and the National Science
Foundation.  This research has made use of NASA's Astrophysics Data
System Bibliographic Services.  The Digitized Sky Surveys were
produced at the Space Telescope Science Institute under
U.S. Government grant NAG W-2166.



\bibliographystyle{apj}
\bibliography{atnf}

\end{document}